\documentclass[twocolumn]{aastex631}

\begin{document}\title{Far-Infrared Luminosity Bursts Trace Mass Accretion onto Protostars}

\author[0000-0002-3747-2496]{William J. Fischer}
\affiliation{Space Telescope Science Institute, 3700 San Martin Drive, Baltimore, MD 21218, USA}

\author[0000-0002-6073-9320]{Cara Battersby}
\affiliation{University of Connecticut, 196A Auditorium Rd Unit 3046, Storrs, CT 06269, USA}

\author[0000-0002-6773-459X]{Doug Johnstone}
\affiliation{NRC Herzberg Astronomy and Astrophysics, 5071 West Saanich Rd, Victoria, BC V9E 2E7, Canada}
\affiliation{Department of Physics and Astronomy, University of Victoria, 3800 Finnerty Rd, Victoria, BC V8P 5C2, Canada}

\author[0000-0002-7482-5078]{Rachel Lee}
\affiliation{University of Connecticut, 196A Auditorium Rd Unit 3046, Storrs, CT 06269, USA}

\author[0000-0003-2248-6032]{Marta Sewi{\l}o}
\affiliation{Exoplanets and Stellar Astrophysics Laboratory, NASA Goddard Space Flight Center, Greenbelt, MD 20771, USA}
\affiliation{Department of Astronomy, University of Maryland, College Park, MD 20742, USA}
\affiliation{Center for Research and Exploration in Space Science and Technology, NASA Goddard Space Flight Center, Greenbelt, MD 20771, USA} 

\author[0000-0002-1700-090X]{Henrik Beuther}
\affiliation{Max Planck Institute for Astronomy, Königstuhl 17, D-69117, Heidelberg, Germany}

\author[0000-0002-9017-3663]{Yasuhiro Hasegawa}
\affiliation{Jet Propulsion Laboratory, California Institute of Technology, Pasadena, CA 91109, USA}

\author[0000-0001-6431-9633]{Adam Ginsburg}
\affiliation{Department of Astronomy, University of Florida, PO Box 112055, USA}

\author[0000-0001-7552-1562]{Klaus Pontoppidan}
\affiliation{Jet Propulsion Laboratory, California Institute of Technology, Pasadena, CA 91109, USA}

\begin{abstract}
Evidence abounds that young stellar objects undergo luminous bursts of intense accretion that are short compared to the time it takes to form a star. It remains unclear how much these events contribute to the main-sequence masses of the stars. We demonstrate the power of time-series far-infrared (far-IR) photometry to answer this question compared to similar observations at shorter and longer wavelengths. We start with model spectral energy distributions that have been fit to 86 Class 0 protostars in the Orion molecular clouds. The protostars sample a broad range of envelope densities, cavity geometries, and viewing angles. We then increase the luminosity of each model by factors of 10, 50, and 100 and assess how these luminosity increases manifest in the form of flux increases over wavelength ranges of interest. We find that the fractional change in the far-IR luminosity during a burst more closely traces the change in the accretion rate than photometric diagnostics at mid-infrared and submillimeter wavelengths. We also show that observations at far-IR and longer wavelengths reliably track accretion changes without confusion from large, variable circumstellar and interstellar extinction that plague studies at shorter wavelengths. We close by discussing the ability of a proposed far-IR surveyor for the 2030s to enable improvements in our understanding of the role of accretion bursts in mass assembly.
\end{abstract}

\section{Introduction}\label{s.intro}

In young stellar objects (YSOs), stochastic enhancements of the accretion rate from the protoplanetary disk onto the star seem to play an important role in building up the stellar mass \citep{kenyon1990,offner2011,dunham2012,fischer2019,lee2021,zakri2022,wang2023}. To some degree, these events supplement the consistent but slowly declining flow of disk mass that persists for a few million years. The stochastic enhancements, known as bursts or outbursts, are observed as dramatic increases in the bolometric luminosities of the objects. 

Historically, these bursts have been divided into two categories. The FU Orionis events (FUors) feature increases of 4 to 5 mag at optical wavelengths over a few months, and this state persists for many years. The first known example began its burst in 1936 \citep{herbig1966} and still has not subsided. The EX Lupi events (sometimes referred to as EXors) feature increases of about 1~mag at optical wavelengths, persist for months to a few years, and may recur in any given star \citep{herbig2007}. In optical and near-infrared (near-IR) spectroscopy, EX~Lup stars resemble young stellar objects with atypically high accretion rates. FU Ori stars, on the other hand, have very different spectra that are dominated by a self-luminous accretion disk. See \citet{reipurth2010}, \citet{audard2014}, Section 2.5 of \citet{hartmann2016}, and \citet{fischer2023} for reviews of the burst phenomenon.

Descriptions of these two classes are necessarily somewhat vague due to the diversity of light curves and spectra seen within each. In recent years, two developments have complicated the picture further. First, many outbursts with intermediate photometric and spectroscopic characteristics have been observed. V1647 Ori has been considered emblematic of these \citep{contreraspena2017}. It began an outburst in 2003, faded, and then recovered to its outburst state in 2008. Spectroscopically, it resembles an FU Ori star except for \ion{H}{1} and CO emission that are more typical of EX Lup stars \citep{connelley2018}.

Second, infrared surveys of star-forming regions or the Galactic plane \citep{scholz2013,carattiogaratti2017,contreraspena2017,lucas2017,fischer2019,lucas2020,park2021,zakri2022} and targeted submillimeter observations \citep{herczeg2017,lee2021} have detected bursts from objects with SEDs consistent with protostars that are still deeply embedded in their natal envelopes. Mid-IR searches frequently define bursts as brightenings in excess of 1 or 2 mag, finding that they occur in a given protostar on $\sim$ 1000 yr timescales \citep{park2021,zakri2022}. These are less dramatic and more frequent than the classical FU Ori bursts, which brighten in the optical by $\sim$ 5 mag and occur on $\sim$ 10,000 yr timescales \citep{hillenbrand2015}. Therefore it is crucial to continue investigating these protostellar bursts to constrain the responsible physics.

\citet{safron2015} reported the first known burst of a Class~0 protostar, in which the infalling envelope is inferred to still contain more mass than the star itself. The burst, of HOPS 383, was discovered in an analysis of 24 \micron\ photometry. This protostar brightened in 2004--2006 and had faded by 2017 \citep{grosso2020}, a light curve suggestive of V1647 Ori. Due to its embedded nature, no optical/near-IR spectroscopy was possible to determine whether it had features similar to EX Lup, FU Ori, or something else.

The case of HOPS 383 is instructive because it tests the limits of what we can determine without spectroscopy and with only sporadic photometry at the necessary wavelengths. While the onset and end of the burst could be dated with reasonable precision, it was difficult to monitor the evolution of the light curve at intermediate times, and it was impossible to grasp the physical insight into the nature of the burst that would have been possible with optical and near-IR spectroscopy.

Due to these observational limitations, it is at present difficult to answer a fundamental question about the assembly of low-mass stars: What fraction of their main-sequence masses is assembled in stochastic bursts? To understand this, we need to determine, at least in a statistical sense, the amplitudes and durations of bursts. Recently, \citet{wang2023} presented a framework for using burst durations and amplitudes to infer what fraction of a star's mass is accreted during each mode and applied it to 70 years of photometry of EX Lup. They concluded that the largest bursts are responsible for about half of the accreted mass during the monitoring time, with the other half arising from the combination of small bursts and quiescent accretion.

It is particularly important to measure the burst durations and amplitudes for Class~0 protostars, before the majority of the stellar mass has been assembled. This paper demonstrates that a far-infrared (far-IR) survey of Galactic star-forming regions is essential for determining these quantities for statistically significant numbers of Class~0 protostars. Protostellar SEDs for a wide range of evolutionary states and viewing angles peak in the far-IR \citep{whitney2003}, so they are most easily detected at such wavelengths.

For the protostellar mass assembly question specifically, far-IR monitoring has two additional benefits. First, this wavelength range gives us the most reliable estimates of the true burst amplitudes. In the far IR, the spectral energy distribution (SED) is most directly responsive to changes in the bolometric luminosity and therefore the accretion rate. Second, the far IR allows us to circumvent the high circumstellar and interstellar extinction that cause incompleteness in mid-infrared (mid-IR) surveys and that can itself be time dependent, producing spurious burst-like signals.

In this work, we start with 1.2--870 \micron\ SEDs that were fit by \citet{furlan2016} to 92 Class~0 protostars, the largest population of this class in a single molecular cloud complex within 500 pc, the Orion Molecular Clouds. This allows us to sample the range of protostellar luminosities and evolutionary states present in a real cloud complex. For each SED, we increase the luminosity in the associated model by fixed factors and consider how effectively those factors are recovered by photometry in different wavelength regimes. In a companion paper, Lee et al.\ (in preparation) consider the observational cadence needed for a far-IR survey to quantify burst rates, amplitudes, and durations.

In Section~\ref{s.modeling}, we introduce the set of model SEDs used to fit the Class~0 Orion protostars and how these models are modified to account for bursts. In Section~\ref{s.bursts} we examine how burst amplitudes depend on wavelength. In Section~\ref{s.extinction} we show the benefits of the far IR for circumventing foreground extinction. In Section~\ref{s.prima} we discuss how future far-IR missions such as the proposed Probe Far-Infrared Mission for Astrophysics (PRIMA) will be useful in addressing the issues presented here. Section~\ref{s.conclusions} presents our conclusions.

\clearpage
\section{SED Modeling}\label{s.modeling}

Here we describe how model SEDs are used to infer the effects of bursts on photometry of Class~0 protostars at different wavelengths. We start with a set of model SEDs used to fit the multiwavelength SEDs of Class~0 Orion protostars by \citet{furlan2016} in the context of the Herschel Orion Protostar Survey (HOPS), a key program of the Herschel Space Observatory. Led by PI S.\ T.\ Megeath, HOPS is described by \citet{fischer2020}. We then modify these SEDs to account for bursts of varying intensity and examine how these bursts are reflected in photometry at different wavelengths.

\subsection{The HOPS Model Grid}

\citet{furlan2016} presented a grid of 30,400 model SEDs designed to infer physical properties of protostars. The grid contains 3040 model protostars, and each has ten SEDs as viewed from different angles. These model SEDs were then fit to the observed SEDs of 319 Orion protostars, constructed with data from the Two-Micron All-Sky Survey (2MASS; \citealt{skrutskie2006}), the Spitzer Space Telescope \citep{werner2004}, the Herschel Space Observatory \citep{pilbratt2010}, and the Atacama Pathfinder Experiment (APEX; \citealt{siringo2009,siringo2010}). Table~\ref{t.data} gives the wavelengths and approximate angular resolutions of datasets used for the SEDs. The techniques used to obtain the photometry are explained in Section 3.1 of \citet{furlan2016}.

\begin{deluxetable}{lccc}
    \tablewidth{0pt}
    \tablecaption{Data Used in Orion SEDs\label{t.data}}
    \tablehead{\colhead{Instrument} & \colhead{Wavelengths (\micron)} & \colhead{Resolution (\arcsec)}}
    \startdata
    2MASS & 1.2, 1.6, 2.2 & 4 \\
    Spitzer/IRAC & 3.6, 4.5, 5.8, 8.0 & 2 \\
    Spitzer/IRS & 5.2--14 & 4 \\
    Spitzer/IRS & 14--38 & 11 \\
    Spitzer/MIPS & 24 & 6 \\
    Herschel/PACS & 70 & 6 \\
    Herschel/PACS & 100 & 7 \\
    Herschel/PACS & 160 & 12 \\
    APEX/SABOCA & 350 & 7 \\
    APEX/LABOCA & 870 & 19 \\
    \enddata
\end{deluxetable}

Of the 319 protostars, \citet{furlan2016} determined 92 to be of Class 0, the least evolved class, via their mid-IR spectral slopes and their bolometric temperatures. In Section~\ref{s.bursts}, we analyze potential bursts from 86 of these Class 0 protostars, dropping six sources where the model SED fits either lack sufficient mid-IR flux for the analysis or are already near the maximum luminosity in the grid.

\begin{figure}
    \includegraphics[width=\columnwidth]{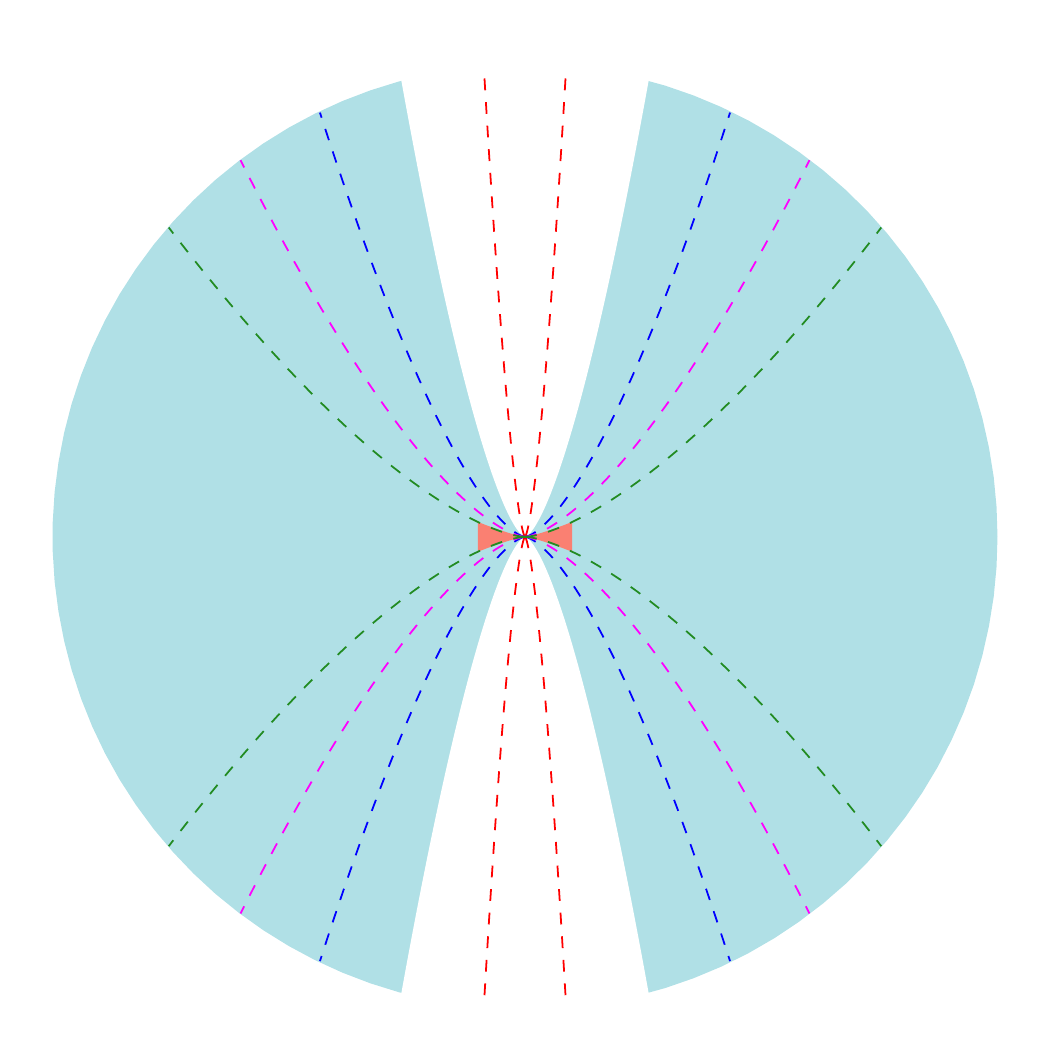}
    \caption{Radiative transfer model geometry. There is a central star, not visible at this scale, and a flared disk (salmon). The radius of the disk can vary. For visibility, here it is shown at twice the size of the largest model. Additionally there is an envelope (light blue) of fixed radius 10,000 au. The envelope density profile follows the \citet{terebey1984} rotating collapse prescription and is normalized to one of 19 values. The envelope cavity has an opening angle that ranges from 5$^\circ$ to 45$^\circ$ in steps of 10$^\circ$. All five possibilities are shown with dotted lines, and the 15$^\circ$ case is shown as evacuated. Finally, the system can be viewed over a range of angles from nearly pole-on (18$^\circ$) to nearly edge-on (87$^\circ$).}\label{f.diagram}
\end{figure}

The grid is described in full detail by \citet{furlan2016}; here we summarize the important properties. Models are calculated with the 2008 version of the HOCHUNK radiative transfer code \citep{whitney2003}. Each model contains a central protostar, a circumstellar disk, and an envelope with radius 10,000 au. The envelope density is that of a rotating, collapsing core with a fixed infall rate \citep{terebey1984}, plus a bipolar cavity evacuated by an outflow. Figure~\ref{f.diagram} illustrates the axisymmetric geometry of the model. 

The disk and envelope consist of dust with opacities from \citet{ormel2011} that account for grain growth at an age of 3 $\times$ 10$^5$ yr and include the effects of ices. Properties relevant to embedded protostars are varied, as described in the next two paragraphs, while others are held constant. Constant parameters of note include the stellar mass (0.5 $M_\odot$), stellar temperature (4000 K), disk mass (0.05 $M_\odot$), and envelope radius (10,000 au); see Table 3 of \citet{furlan2016} for the full list. This is an important distinction from larger, more widely used grids that vary stellar properties and other features that have little effect on SEDs when there is a dense envelope present.

Four parameters are varied in the grid. First, the total luminosity of the protostar can take on one of eight values extending from 0.1 $L_\sun$ to 303 $L_\sun$ in roughly equal logarithmic steps. Because the stellar and accretion luminosity are both reprocessed by the disk and envelope, protostellar SEDs depend only weakly on the relative contributions of these inputs. Therefore the luminosity is modified by adjusting the radius of the star or the accretion rate onto the star. To improve the fits by allowing for a continuous range of luminosities, a limited-range scaling factor is applied, based on the finding that the shape of an SED is only weakly dependent on small changes in the luminosity \citep{kenyon1993}. Table~\ref{t.luminosities} shows how the required luminosities are obtained for different choices of stellar radius, accretion rate, and scaling factor.

Second, the envelope density can take one of 19 values. The envelope density at 1000 au can be 0 or can range from $1.19 \times 10^{-20}$ g cm$^{-3}$ to $1.78 \times 10^{-16}$ g cm$^{-3}$ in roughly equal logarithmic steps. The envelope density elsewhere is set by the rotating collapse model mentioned above. Third, the outer radius of the disk, which is also the centrifugal radius of the envelope, can be 5, 50, 100, or 500 au. Fourth, the cavity opening angle can be 5, 15, 25, 35, or 45$^\circ$.

To account for inclination ($i$) dependence, each model can be viewed from one of ten angles ranging from 18$^\circ$ (almost face-on) to 87$^\circ$ (almost edge-on) in equal steps of $\cos i$. Model fluxes are determined for 24 apertures that run from 420 au to 10,080 au in steps of 420 au. For each wavelength, an aperture is chosen that represents the angular resolution of the relevant instrument (Table~\ref{t.data}), assuming a source distance of 420 au. Finally, foreground extinction is applied with laws from \citet{mathis1990} if $A_J < 0.76$ or \citet{mcclure2009} otherwise. For each protostar, the best model is the one that has the least average weighted logarithmic deviation between the observed and model SED \citep{fischer2012}. 

\subsection{Adding Bursts to the HOPS Model Grid}\label{s.adding}

\begin{deluxetable}{lccc}
    \setlength{\tabcolsep}{10pt}
    \tablewidth{0pt}
    \tablecaption{Dependence of Luminosity on Model Parameters\label{t.luminosities}}
    \tablehead{\colhead{$L$} & \colhead{$R_*$} & \colhead{$\dot{M}$} & \colhead{Scaling Factor$^1$} \\[-0.2cm] \colhead{($L_\sun$)} & \colhead{($R_\sun$)} & \colhead{($M_\sun$ yr$^{-1}$)} & \colhead{}}
    \startdata
     0 -- 0.2 & 0.67 & 0 & 0 -- 2.0 \\
     0.2 -- 0.6 & 0.67 & $1.14\times10^{-8}$ & 0.67 -- 2.0 \\
     0.6 -- 2.0 & 0.67 & $5.17\times10^{-8}$ & 0.60 -- 2.0 \\
     2.0 -- 6.2 & 2.09 & $3.67\times10^{-7}$ & 0.65 -- 2.0 \\
     6.2 -- 20.2 & 2.09 & $1.63\times10^{-6}$ & 0.61 -- 2.0 \\
     20.2 -- 60.4 & 6.61 & $1.14\times10^{-5}$ & 0.67 -- 2.0 \\
     60.4 -- 202 & 6.61 & $5.15\times10^{-5}$ & 0.60 -- 2.0 \\
     202 --  & 6.61 & $1.66\times10^{-4}$ & 0.67 --
    \enddata
    \vspace{0.2cm}
    $^1$The scaling factor allows a continuous distribution of protostellar luminosities to be fit with a discrete grid of models; see the discussion in Section~\ref{s.adding}. The scaling factor for the most luminous model could, in principle, be arbitrarily large to accommodate arbitrarily large luminosities.
\end{deluxetable}

\begin{figure*}
    \includegraphics[width=0.49\textwidth]{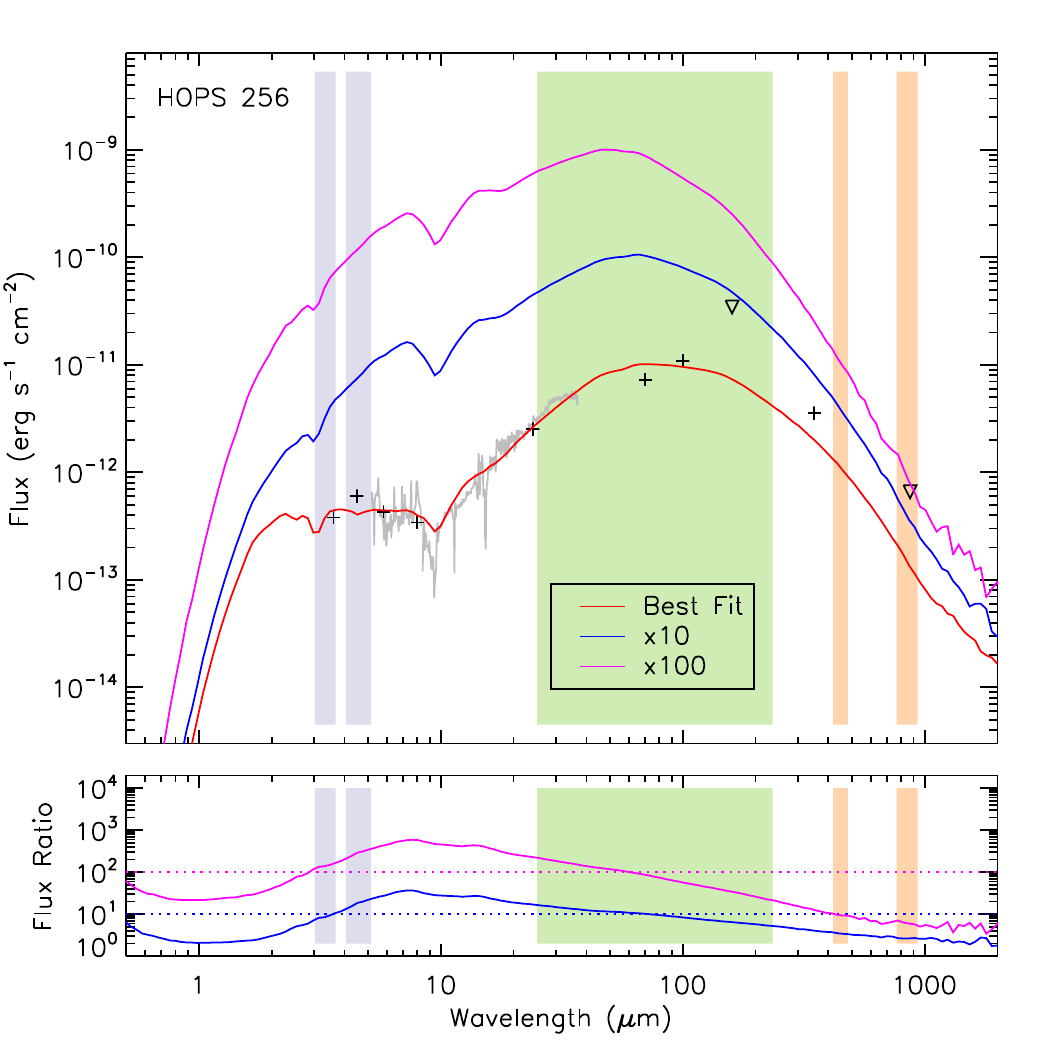}
    \includegraphics[width=0.49\textwidth]{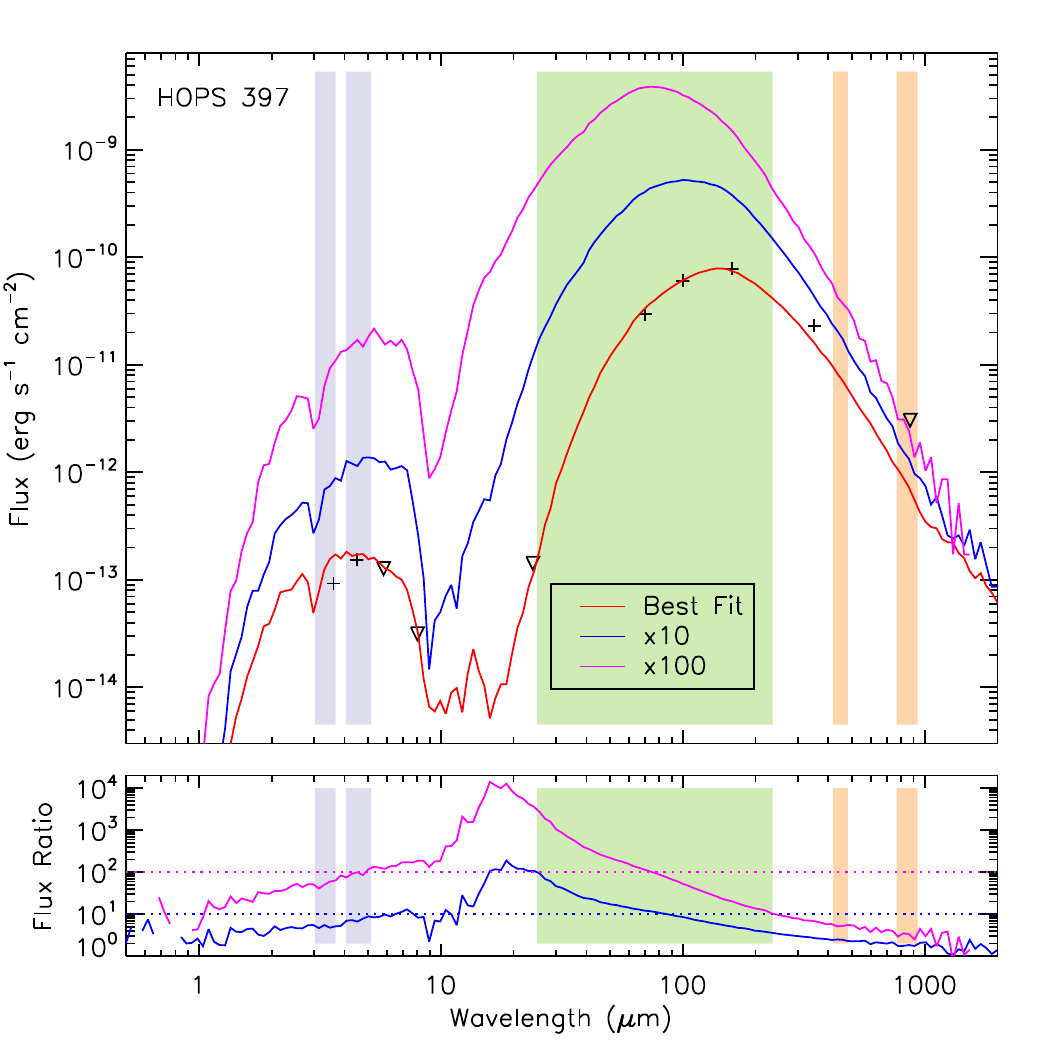}
    \caption{Demonstration of the burst procedure for two Class~0 protostars. HOPS 256 (left) was modeled by \citet{furlan2016} with a nearly face-on viewing angle of 18$^\circ$, and HOPS 397 (right) was modeled by the same authors with a nearly edge-on viewing angle of 87$^\circ$. In each upper panel, black symbols mark photometry ($+$ signs) or limits (triangles) from Spitzer, Herschel, and APEX. The gray spectrum for HOPS 256 is from Spitzer. The red curve is the best-fit SED from a radiative transfer model, and the blue and magenta curves show the SEDs for the same models, scaled up by factors of 10 or 100 in luminosity, respectively. Rectangles show the wavelength regions of photometry discussed in Section~\ref{s.bursts}: purple for the WISE 3.4 and 4.6 \micron\ bands, green for one of the far-IR space mission concepts (see Section~\ref{s.prima}), and orange for the JCMT 450 and 850~\micron\ bands. In each lower panel, the ratios of the factor-of-10 and factor-of-100 burst SEDs to the best-fit SEDs are plotted with solid blue and magenta lines, while dotted blue and magenta lines mark fixed ratios of 10 and 100. In the mid IR, the observed flux changes are greater than the luminosity changes for the face-on protostar and less than the luminosity changes for the edge-on protostar. In the far IR, the integrated flux change gives a robust estimate of the luminosity change. In the submillimeter, the flux changes are significantly less than the luminosity changes.}\label{f.seds}
\end{figure*}

The best-fit model for each protostar has a luminosity associated with it; see Column 8 of Table 1 in \citet{furlan2016}. To simulate bursts, we increase each of these luminosities by factors of 10, 50, and 100 and choose a model from the existing grid with this luminosity, holding other parameters (envelope density, disk radius, cavity opening angle, inclination, and foreground extinction) fixed. Within each of the luminosity ranges shown in Table~\ref{t.luminosities}, the given stellar radius and mass accretion rate are chosen, and a scaling factor within the given range is chosen to reproduce the burst luminosity. As with the initial fitting exercise, we again assume that the SED does not depend significantly on the division between stellar and accretion luminosity or on luminosity changes covered by the range of scaling factors. These assumptions were validated by \citet{furlan2016}.

Figure~\ref{f.seds} shows the original, factor-of-10 burst, and factor-of-100 burst models for the nearly face-on protostar HOPS 256 and the nearly edge-on protostar HOPS 397. The upper panels show how the SEDs change, brightening and shifting to bluer wavelengths as the bursts intensify. The lower panels show how the ratios of the two burst SEDs to their quiescent counterpart vary with wavelength in different ways depending on the details of the best-fit model.

\section{Burst Amplitudes with Wavelength}\label{s.bursts}

\begin{figure*}
    \includegraphics[width=\textwidth]{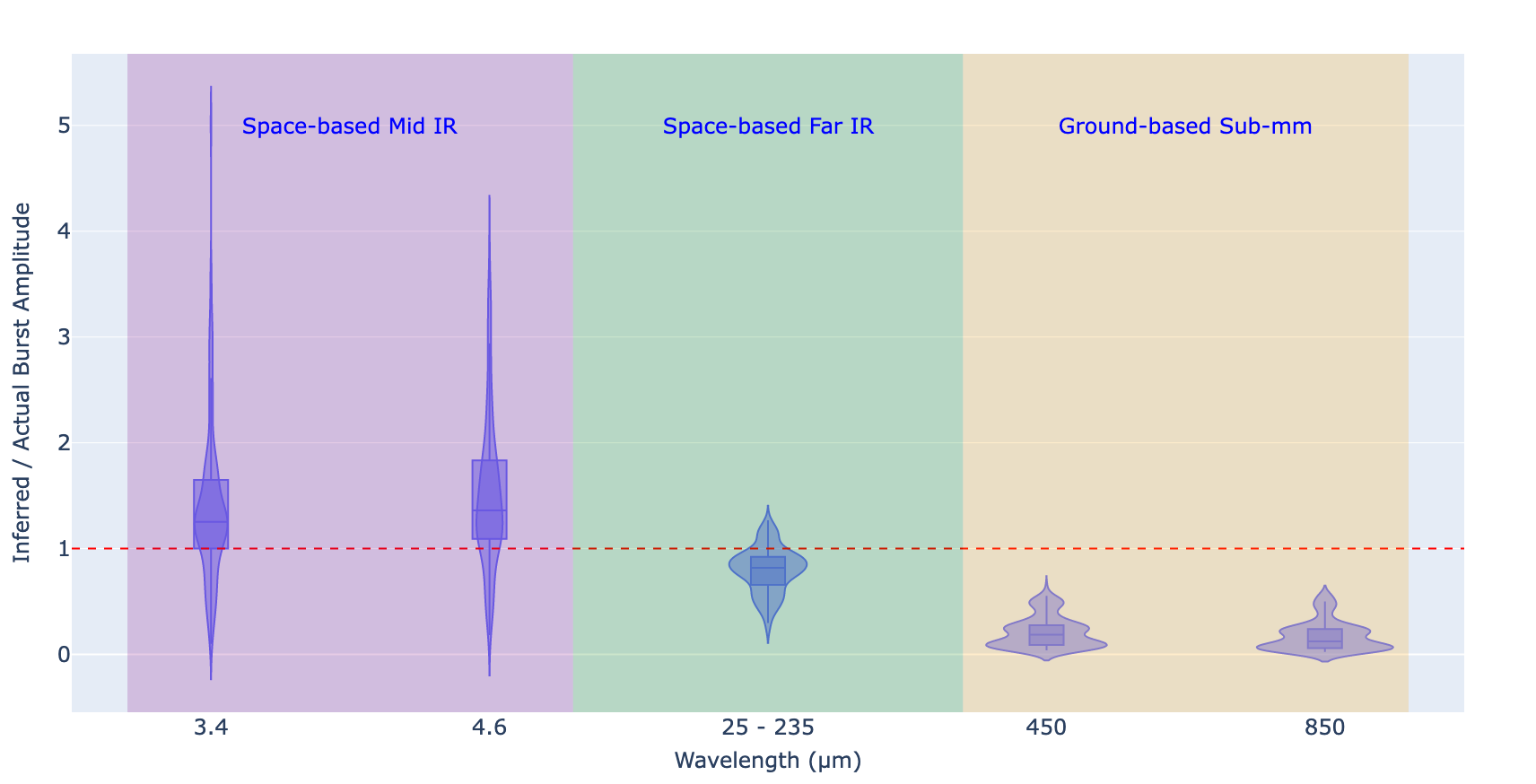}
    \caption{Ratio of inferred to actual burst amplitude for bursts in 86 Class~0 protostars. The 25--235 \micron\ point refers to the result of integrating the SED over far-IR photometric bands from 25 to 235 \micron. In these violin plots,  the shape along the vertical axis shows the continuous distribution of points obtained via kernel density estimation from the discrete measurements. Horizontal lines mark the first quartile (bottom), median (central), and third quartile (top) of each distribution.}\label{f.bursts}
\end{figure*}

For each model, we calculate flux densities (for single wavelengths) or integrated luminosities (for a range of adjacent, simultaneously observed wavelengths) that would be observed by several existing and planned telescopes. To simulate WISE \citep{wright2010} or NEOWISE \citep{mainzer2014} observations, we calculate the 3.4 and 4.6 \micron\ flux densities. To simulate observations such as those obtained by the James Clerk Maxwell Telescope transient survey \citep{herczeg2017}, we calculate the 450 and 850 \micron\ flux densities. Finally, we calculate the far-IR luminosity obtained by integrating under the 25 to 235 \micron\ SED as observed with 16 separate bandpasses placed over this range at roughly equal logarithmic intervals.

We define the inferred burst amplitude in each wavelength range as the flux density (in the mid IR and submillimeter) or luminosity (in the far IR) observed in the burst SED to that of the pre-burst, quiescent SED. We define the actual burst amplitude as the change in the true luminosity of the protostar; i.e., a factor of 10, 50, or 100. We include 86 Class 0 protostars in the sample. Four of the initial 92 are excluded because their pre-burst luminosities exceed 202 $L_\odot$. This places them in the bottom row of Table~\ref{t.luminosities}; in this case, simulating a burst involves multiplying the SED by a constant, which is not instructive. Two additional protostars are excluded because their pre-burst models have no detectable 3.4 \micron\ flux, leaving their post-to-pre-burst ratios poorly constrained.

Figure~\ref{f.bursts} shows how the ratio of the inferred burst amplitude to the actual burst amplitude depends on wavelength. For example, if the model luminosity increased by a factor of 10 but the 3.4 \micron\ flux increased by a factor of 12, the value on the vertical axis of the figure would be 1.2.

In the mid IR, the ratio covers a broad range and is centered slightly above unity. Mid-IR fluxes are heavily dependent on geometry. In protostars that are viewed through their outflow cavities, a disproportionately large amount of the mid-IR flux is emitted along the line of sight, and the fractional increase in the mid-IR flux is large compared to the fractional increase in luminosity. In protostars that are viewed through their disks, very little flux reaches the line of sight, and the opposite occurs. Furthermore, when the envelope is heated by the burst, the location of the effective surface seen by the observer changes and the local temperature associated with this surface also changes such that the emission can vary considerably. For the majority of the envelope, the mid IR is on the Wien side of the SED, so small changes in optical depth or temperature can lead to large changes in flux density. The potential for large changes makes the mid-IR a reasonable wavelength range in which to search for bursts, even if the change in flux density there is likely to overestimate the change in accretion rate.

At submillimeter wavelengths, the SED responds to the temperature change in the disk and envelope, the magnitude of which is significantly reduced compared to the luminosity change \citep{johnstone2013}.

The far IR traces the luminosity change best, in that it is centered just below unity and has a small dispersion. This is for two reasons. First, these changes are determined by integrating over a broad range of wavelengths, from 25 to 235 \micron, capturing the peak of the SED for a broad range of evolutionary states \citep{whitney2003}. Second, the far IR probes luminosity in a geometry-independent way, responding to the heating of the inner envelope, which is fairly uniform regardless of geometry and viewing angle.

An additional advantage to observing with a large number of far-IR bandpasses is that flux ratios among these bandpasses change as outbursts propagate through and heat the protostellar envelope. Figure~\ref{f.seds} shows how the SED peak shifts to shorter wavelengths as the bursts intensify. While any single bandpass in this range would reflect the actual burst amplitude reasonably well, integrating over several bandpasses results in a more robust measurement.

The choice in the far IR of 16 separate bandpasses extending from 25 to 235 \micron\ is motivated by plans for a proposed space mission concept (see Section~\ref{s.prima}). To estimate the extent to which the result depends on the specific number of bandpasses and their wavelengths, we repeated the exercise for two previous far-IR instruments. The Photodetector Array Camera and Spectrometer (PACS; \citealt{poglitsch2010}) aboard the Herschel Space Observatory had three bandpasses centered at 70, 100, and 160 \micron. The Far-Infrared Surveyor (FIS; \citealt{kawada2007}) aboard the AKARI space mission \citep{murakami2007} had four bandpasses centered at 65, 90, 140, and 160 \micron.

\begin{figure*}
    \includegraphics[width=\textwidth]{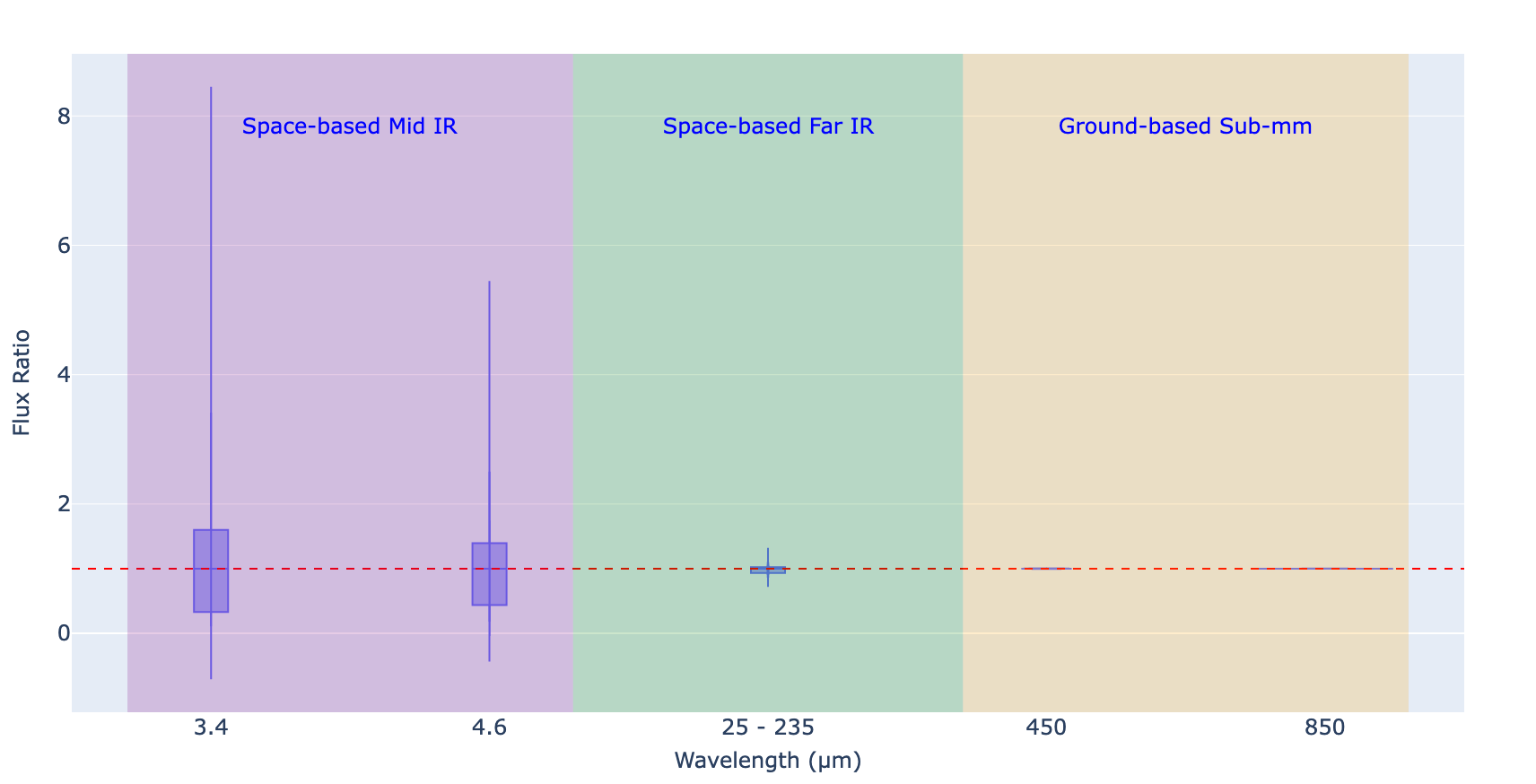}
    \caption{Flux ratios for the ensemble of protostars if the $V$ band extinction $A_V$ toward each protostar is changed by a randomly selected value between $-$30 and 30 mag, enforcing a minimum $A_V$ of 0. The submillimeter ratios are uniformly too close to unity to be discerned well in the plot.}\label{f.bursts_av}
\end{figure*}

The far-IR symbol in Figure 3 has a median of 0.82, and the central half of the distribution spans a range of 0.26. The ratios for the Herschel simulation have a median of 0.68, and the central half of the distribution spans a range of 0.31. The ratios for the AKARI simulation have a median of 0.84, and the central half of the distribution spans a range of 0.39. The results are similar for these three collections of far-IR bandpasses, but the actual burst amplitudes are more closely recovered, with a smaller dispersion, when many bandpasses are available across the full far-IR range.

These theoretical results are similar to those found by \citet{macfarlane2019} and \citet{baek2020}, who used similar methods except that we present the effect of bursts on SEDs that have been fit to an ensemble of 86 well-characterized Class~0 protostars in a single molecular cloud complex. Similar findings regarding the wavelength dependence of flux changes were reported by \citet{stecklum2021} in an analysis of the burst of G353.93$-$0.03, a massive YSO. Our results also build on the mid-IR versus submillimeter multiwavelength analysis of observed protostellar variability by \citet{contreraspena2020}. That study revealed both a clear correlation in brightness variations between the two widely separated wavelengths and a clear indication that neither wavelength responded in a manner directly proportional to the changing central source luminosity.

\section{Avoiding Extinction with Far-IR Surveys}\label{s.extinction}

Another reason to survey for bursts in the far IR is to avoid the complications of large, variable circumstellar and interstellar extinction that affect observations at shorter wavelengths. This causes three distinct problems in assessing the rates, amplitudes, and durations of accretion bursts.

First, it can be ambiguous whether brightness changes in the mid IR are due to accretion or to changing foreground extinction. Redder mid-IR colors during outburst can be used to rule out a dust clearing event, since this would yield bluer colors (e.g., \citealt{zakri2022}), but the interpretation is more difficult in the event of bluer mid-IR colors. The far IR is only mildly sensitive to changes in extinction, making it a more robust probe.

We demonstrate the robustness of far-IR and submillimeter photometry against variable extinction in Figure~\ref{f.bursts_av}. For each of the 86 Class 0 protostars, we consider the foreground $A_V$ modeled by \citet{furlan2016} and change this by a randomly selected value between $-$30 and 30 mag, enforcing a minimum $A_V$ of 0. The threshold of 30 is based on the maximum change seen in V2492 Cyg, a star with a history of extreme variability due to both accretion and extinction changes \citep{hillenbrand2013}. The mid-IR fluxes vary by a large range, while the far-IR fluxes vary only slightly, and the variation of the submillimeter fluxes is imperceptible. The distributions are peaked slightly below 1, since increasing $A_V$ by up to 30 is always possible, while a decrease in $A_V$ is capped by our prohibition against negative $A_V$.

Second, point sources detected in mid-IR imaging surveys are often knots of shock emission \citep{gutermuth2009,koenig2014} or scattering surfaces \citep{yoon2022} within outflows. These may be several arcseconds from the protostar, while the protostar itself is hidden behind intervening dust. Such locations could be directly illuminated by an accretion burst in the central protostar or indirectly heated through shocks associated with varying conditions within the jet-like outflow. Either of these events may occur significantly after the onset of the burst, complicating efforts to understand a given protostar's accretion history.

Third, many Class~0 protostars are too deeply embedded to be detected before a luminosity outburst. The classification of HOPS 383 as a protostar was tenuous before its outburst \citep{safron2015}. Of the 92 Class~0 protostars in Orion \citep{furlan2016}, 16 (17\%) were not classified as protostars based on mid-IR Spitzer data alone and required Herschel far-IR photometry to be classified as such \citep{stutz2013}. In regions that are less well characterized than Orion, far-IR surveys will be needed to recover the full population of young Class~0 protostars.

\section{Prospects for Far-IR Monitoring of Protostellar Outbursts }\label{s.prima}

Far-IR monitoring of enough protostars to understand the role of luminosity bursts in stellar mass assembly will require a new mission with the requisite sensitivity, mapping speed, and wavelength coverage. The Astro2020 Decadal Report recommended ``a far-IR imaging or spectroscopy probe mission,'' prompting NASA to issue an Announcement of Opportunity for an Astrophysics Probe Explorer in July 2023. One of the mission concepts developed in response to this call is the Probe Far-Infrared Mission for Astrophysics (PRIMA; \citealt{glenn2023}), with science instruments well suited for far-IR monitoring of protostellar outbursts.

PRIMA is planned to be a cryogenically cooled far-IR observatory for the 2030s, with a nominal mission lifetime of five years. It will improve mapping speeds by up to four orders of magnitude with respect to its far-IR predecessors and is expected to open wide discovery space.

The two science instruments planned for PRIMA are PRIMAger, an imager \citep{burgarella2023}, and FIRESS, a spectrometer \citep{bradford2023}. PRIMA\-ger is a sensitive multi-band spectrophotometric imager. It offers hyperspectral narrow-band imaging (with spectral resolving power $R \sim 10$) from 25 to 80~\micron\ and polarimetric capabilities in four broadband filters from 80 to 260~\micron. FIRESS offers coverage from 24 to 235 \micron, in either low spectral resolution ($R \sim 130$) or moderate spectral resolution ($\gtrsim$ a few $\times$ 1000).

The PRIMA instruments are well suited to monitoring protostellar SEDs over time and performing follow-up observations of spectral lines to investigate changes in chemistry and physical properties during accretion events. The wide bandwidth of PRIMAger makes it highly effective for monitoring protostellar SEDs, and surveying can be efficient since protostars are highly clustered in the sky. FIRESS can be used to monitor changing physical conditions via analysis of far-IR spectral lines including rotational transitions of H$_2$O, high-J CO, OH, [O {\sc i}] and [C {\sc ii}] (e.g., \citealt{manoj2013}; \citealt{karska2018}).

Lee et al.\ (in preparation) argue that a monitoring survey of 2000 protostars is necessary and sufficient to robustly answer the question of whether protostars gain the majority of their mass through accretion bursts or through steady-state processes. They discuss monitoring 2000 protostellar SEDs with PRIMA, tracking their brightness over the five-year mission, and comparing the fluxes to those obtained twenty years earlier with Herschel. An initial spectral scan of all protostars with FIRESS, followed by triggered FIRESS observations during and after observed bursts, will enable detailed monitoring of the changing physical conditions. These observations are also critical for measuring protostellar burst light curves to constrain both the burst mechanisms and the event durations.

\section{Conclusions}\label{s.conclusions}

With model SEDs that were fit to the observed SEDs of 86 Class~0 protostars in the Orion molecular clouds by \citet{furlan2016}, we explored how major luminosity outbursts due to accretion manifest as a function of wavelength. The protostars sample the ranges of envelope density, cavity opening angle, and viewing angle encountered in the largest star-forming region in the neareast 500 pc.

We find that photometry over several distinct bands in the far-IR range, e.g., 25 to 235 \micron, is essential for three reasons. First, the change in luminosity seen over this range, measured as a fraction of the total luminosity change, has a small dispersion and is close to unity, unlike that seen at shorter or longer wavelengths. Second, the location of the SED peak shifts to shorter wavelengths during the burst, and this can be tracked with multiple bandpasses over this wavelength range. Third, this wavelength range is less susceptible to the confusing effects of variable circumstellar and interstellar extinction than shorter wavelengths.

Due to these advantages, a far-IR time-domain survey of nearby star-forming regions as outlined in Lee et al.\ (in preparation) may lead to a breakthrough in our understanding of the importance of outbursts in assembling the main-sequence masses of stars.

\pagebreak
We thank Bringfried Stecklum and Lynne Hillenbrand for feedback on a draft of this work. M.S. acknowledges support from the NASA ADAP Grant Number 80NSSC22K0168. The material is based upon work supported by NASA under award number 80GSFC21M0002 (M.S.).\ D.J. is supported by NRC Canada and by an NSERC Discovery Grant. C.B. gratefully acknowledges funding from the National Science Foundation under Award Nos.\ 1816715, 2108938, 2206510, and CAREER 2145689, as well as from the National Aeronautics and Space Administration through the Astrophysics Data Analysis Program under Award No.\ 21-ADAP21-0179 and through the SOFIA archival research program under Award No.\ 09$\_$0540. This research was carried out in part at the Jet Propulsion Laboratory, California Institute of Technology, under a contract with the National Aeronautics and Space Administration (80NM0018D0004).

\bibliography{prima}{}
\bibliographystyle{aasjournal}

\end{document}